\documentclass{article}
\usepackage{amsmath}
\usepackage{amsfonts}
\usepackage{amssymb}
\usepackage{pstricks,pst-node,amssymb,amsmath,graphics,latexsym,tabularx,shapepar}
\usepackage{amsthm}
\usepackage{graphicx}
\usepackage{float}
\newtheorem{theorem}{Theorem}[section]

\newtheorem{corollary}[theorem]{Corollary}

\newtheorem{lemma}[theorem]{Lemma}

\newtheorem{proposition}[theorem]{Proposition}
\newtheorem{remark}[theorem]{Remark}

\newcommand{\bpartial}{\mathop{\partial\kern -4pt\raisebox{.8pt}{$|$}}}
\newcommand{\bra}{\mathopen{[\kern-1.6pt[}}
\newcommand{\ket}{\mathclose{]\kern-1.5pt]}}
\newcommand{\bbra}{\mathopen{[\kern-2.2pt[\kern-2.3pt[}}
\newcommand{\bket}{\mathclose{]\kern-2.1pt]\kern-2.3pt]}}

\renewcommand{\and}{ and }
\makeindex

\begin{document}


\title{Lie symmetry analysis and similarity reductions for the tempered-fractional Keller--Segel system}
\author{
Ghorbanali Haghighatdoost\thanks{Corresponding author} \and Mustafa Bazghandi \\
\textit{\small Department of Mathematics, Azarbaijan Shahid Madani University, Tabriz, Iran} \\
\small E-mails: gorbanali@azaruniv.ac.ir, mostafabazghandi2001@gmail.com
}

\maketitle

\begin{abstract}
We study the tempered-fractional Keller--Segel (TFKS) model of chemotaxis with anomalous diffusion. The short-jump diffusion approximation is employed to simplify the tempered fractional Laplacian into an effective local diffusion operator, while a gauge transformation converts the tempered Caputo derivative into the standard Caputo form, thereby enabling the application of Lie symmetry analysis. The resulting reduced TFKS system is then analyzed by deriving its complete set of Lie point symmetries and constructing an optimal system of one-dimensional subalgebras. This yields similarity reductions to ordinary differential equations (ODEs) and exact analytical solutions of the reduced system. These results offer insights into the long-term behavior and aggregation dynamics of the TFKS model and present a methodology applicable to other tempered fractional differential equations.
\end{abstract}
\noindent {\bf Keywords}: Lie symmetries, Similarity solutions, Keller--Segel model, Tempered fractional differential equations.

\noindent{\bf AMS}:35R11, 70G65, 76M60.

\section{Introduction}
Mathematical modeling has become an indispensable tool for analyzing biological systems, providing valuable insights into mechanisms that are often too complex to fully understand through experimental approaches alone. One area where mathematical analysis has played a crucial role is in the study of \textit{chemotaxis}, the directed movement of motile organisms, such as bacteria, cells, or particles, in response to chemical gradients. Chemotaxis is fundamental to a variety of biological processes, including embryonic development, wound healing, immune responses, and the formation of bacterial colonies. Chemotactic movement often follows a  L\'{e}vy-like pattern, characterized by long jumps and heavy-tailed step lengths. Due to its prevalence in both nature and medicine, understanding the mechanisms underlying chemotactic movement has long been a central focus of mathematical biology.

A seminal framework for modeling chemotaxis was proposed in the 1970s by Evelyn Keller and Lee Segel, now widely known as the classical Keller--Segel (KS) model \cite{KellerSegel1970}. 
This model has become the cornerstone of chemotaxis theory, successfully capturing essential features of population aggregation driven by chemical cues. The KS system has been extensively studied, both analytically and numerically, and has inspired a wide range of theoretical research (e.g., \cite{Osaki2001,Ryu2001}). While the KS model has been foundational, it has several limitations. Specifically, it assumes normal diffusion, which makes it incapable of adequately modeling anomalous diffusion with finite propagation speed. Additionally, it struggles to account for long-range effects in a controlled manner and is well known for exhibiting finite-time blow-up under certain conditions (i.e., cell density collapsing into a singularity), which may not always be biologically realistic.

The fractional Keller--Segel equation (FKS) extends the classical model by incorporating anomalous diffusion and non-local transport, offering greater flexibility in capturing realistic movement patterns, preventing overly singular aggregation, and aligning more closely with experimental observations of biological motility. The mathematical properties of the FKS equation have been the focus of extensive recent research (e.g., \cite{Chen2025,Gao2025,Kumar2017}). However, like the classical KS model, the FKS may still exhibit finite-time blow-up under certain conditions.
\par 
The tempered fractional Keller--Segel (TFKS) model has been introduced as a refinement of the classical KS model, incorporating tempered fractional operators to better capture the realistic features of chemotactic movement \cite{Zhang2023, Zhang2024}. The TFKS model accounts for anomalous diffusion while tempering it, introducing an exponential tempering function that cuts off extreme long jumps while preserving the memory and non-local nature of the movement. The TFKS model considered in this work extends this formulation by additionally incorporating a logistic growth term. The exponential tempering mechanism governs the long-time behavior of chemotactic cells by inducing a gradual monotonic decay in cell density, thereby preventing unrealistic persistence and yielding biologically plausible dynamics. In addition, the logistic growth term regulates population dynamics by limiting cell proliferation, reducing the tendency toward unbounded aggregation and, under suitable conditions, preventing finite-time blow-up \cite{Zhuang2019}. Overall, the TFKS model improves upon the classical KS model by combining the ability to describe both anomalous diffusion and long-range chemotactic movement with biologically realistic constraints. It also provides a flexible bridge between classical and fractional models.

\par 
The aim of this paper is to study TFKS system with short-jump diffusion for $0<\alpha< 1$ in a one spatial dimension. We carry out a Lie symmetry analysis of the reduced TFKS system, which is obtained by applying a gauge transformation and the short-jump diffusion approximation. We subsequently reduce the system to ODEs and derive exact solutions of the reduced local TFKS system. To the best of our knowledge, the TFKS system has not yet been studied using Lie symmetry groups.

\par 
The Lie symmetry method is one of the most effective techniques for obtaining similarity solutions of
differential equations. Moreover, by examining the symmetries of a system, one can investigate various properties
of partial differential equations (PDEs), such as differential invariants, group classification, optimal systems,
exact solutions, and conservation laws. The Lie method is well-documented in several monographs \cite{bluman2010applications,olver2000applications} and has been widely applied in numerous studies (e.g. \cite{Bazghandi2019,Haghighatdoost2023,Haghighatdoost2025,lashkarian2017}).

\par
This paper is organized as follows: In Section \ref{sec:priliminaries}, we introduce a gauge transformation to convert the tempered Caputo derivative into the standard Caputo form and employ the short-jump diffusion approximation to replace the tempered fractional Laplacian with a local surrogate, thereby enabling the use of classical Lie point symmetry methods. In Section \ref{sec:lie_symmetries}, we apply this methodology to the tempered-fractional Keller--Segel (TFKS) system and derive its Lie point symmetries. In Section \ref{sec:OptimalSys}, we find the optimal system for one-dimensional Lie algebras. In Section \ref{sec:SimilaritySolutions}, we obtain similarity reductions of the reduced local TFKS system. Finally, in Section \ref{sec:ExactSolutions}, we obtain the exact solutions of the reduced systems and discussing their biological significance.

\section{Preliminaries}\label{sec:priliminaries}
In this section, we outline a practical methodology for extending Lie symmetry analysis to tempered fractional PDEs. The Lie symmetry method fundamentally relies on local prolongation formulas, whereas fractional derivatives are inherently non-local, which complicates the direct application of standard Lie techniques. Although Lie methods have been extended to a variety of fractional PDEs over the past two decades (e.g. \cite{Cheviakov2017,Gazizov2009,Liu2024,Soares2024,Yu2024} ), the presence of tempering--which multiplies the integral kernel by an exponential factor--further complicates both the invariance condition and the appropriate prolongation formulas.
\par 
We consider the following class of tempered fractional evolution equations:
\begin{equation*}
{}^{C}D_t^{\alpha , \lambda} u(x,t)=\mathcal{F}\left( x,t,u,\nabla u,\mathcal{L}^{(\lambda)}_\alpha u \right),
\end{equation*}
where ${}^{C}D_t^{\alpha , \lambda}$ denotes the tempered Caputo fractional derivative of order $0<\alpha\le1$, $\mathcal{L}_{\alpha}^{(\lambda)}$ denotes the tempered fractional Laplacian, and $\mathcal{F}$ is a nonlinear operator that may contain reaction, convection, and diffusion terms.
\par 
The principal mathematical difficulty in analyzing such equations arises from the simultaneous presence of the tempered Caputo time derivative and the nonlocal tempered fractional Laplacian. To facilitate analytical treatment, we employ two complementary techniques: a gauge transformation and the short-jump diffusion approximation.
\subsection{Gauge transformation}
Introducing the gauge transformation
\begin{equation*}
v(x,t)=e^{\lambda t}u(x,t),
\end{equation*}
Using the well-known identity for tempered derivatives (see, e.g., \cite{ortigueira2021,Sabzikar2015}):
\begin{equation}\label{eq:identity}
D_t^{\alpha,\lambda}u(t)
= e^{-\lambda t}\,\,{}^{C}D_t^{\alpha}\big(e^{\lambda t}u(t)\big),
\end{equation}
the tempered Caputo derivative is transformed into the standard Caputo derivative,
\begin{equation*}
D_t^{\alpha,\lambda}u
= e^{-\lambda t}\,\,{}^{C}D_t^{\alpha}\big(v\big),
\end{equation*}
Consequently, the temporal operator assumes a form compatible with the existing fractional Lie symmetry framework.
\par 
However, the transformed equation generally still contains the nonlocal tempered fractional Laplacian, which prevents the direct application of classical Lie symmetry methods.
\subsection{Short-jump diffusion approximation}
We adopt the short-jump diffusion approximation to reduce the nonlocal tempered fractional Laplacian to an equivalent local diffusion operator. In the short-jump regime, the nonlocal tempered fractional Laplacian admits a local second-order approximation, which has been widely employed in the analysis of tempered diffusion processes (e.g. \cite{Baeumer2010,Zhang2018}). 
\begin{lemma}\label{lemma:1} 
Let $u(x,t)$ be a sufficiently smooth function, and let $\mathcal{L}_{\alpha}^{(\lambda)}$ denote the tempered fractional Laplacian of order $0<\alpha<2$ with tempering parameter $\lambda>0$. Assume that the characteristic jump length of the underlying L\'evy process is small compared with the macroscopic spatial scale of $u$. Then, under the short-jump diffusion approximation, the tempered fractional Laplacian admits the local asymptotic expansion
\begin{equation*}
 \mathcal{L}_{\alpha}^{(\lambda)}u = c_{2}(\alpha,\lambda)\Delta u +\mathcal{O}(\varepsilon^{2}), \label{eq:shortjump} \end{equation*}
 where $c_{2}(\alpha,\lambda)>0$ denotes the finite second moment of the tempered jump kernel. 
 \par Consequently, the leading-order approximation is
 \begin{equation*} 
 \mathcal{L}_{\alpha}^{(\lambda)}u \approx c_{2}(\alpha,\lambda)\Delta u , 
 \end{equation*}
 \end{lemma}
\begin{proof}
The proof follows from the second-order Taylor expansion of the nonlocal integral representation of the tempered fractional Laplacian. Under the short-jump assumption, odd-order moments vanish because of the symmetry of the tempered L\'evy kernel, while the second-order moment remains finite owing to the exponential tempering (see, e.g. \cite{Baeumer2010,Sabzikar2015,Zhang2018}). Retaining terms up to second order yields the stated approximation, with the neglected higher-order terms being of order $\mathcal{O}(\varepsilon^2)$. 
\end{proof}
\par 
This approximation converts the original nonlocal integro-differential equation into a local fractional partial differential equation while preserving the leading-order effects of tempered anomalous diffusion. As a result, classical prolongation formulas, determining equations, and similarity reduction techniques become directly applicable.
\par 
The analytical framework developed above is not restricted to a particular model but applies to tempered fractional evolution equations whose temporal and spatial operators admit the above transformations. While this approach proves effective, it may obscure the natural scaling relationship between the variables and the tempering parameter. In the next section, this framework is specialized to the tempered-fractional Keller--Segel system, where the gauge transformation and the short-jump diffusion approximation yield a reduced local model suitable for Lie symmetry analysis.
\section{Lie Symmetries of the Reduced TFKS System}\label{sec:lie_symmetries}
\par 
We consider the TFKS system:

\begin{equation}\label{eq:TFKS}
\begin{cases}
&D_t^{\alpha,\lambda}u = D\,\mathcal{L}^{(\lambda)}_{\alpha}\,u \;-\; \chi\,\nabla\!\cdot\!\big(u\,\nabla c\big)\;+\; r\,u\!\left(1-\frac{u}{K_0}\right),\\[2mm]
&\tau_c\,c_t = D_c\,\Delta c \;-\; \kappa\,c \;+\; u, 
\end{cases}
\end{equation}
with coefficients $0<\alpha<2$, $D,D_c,\chi,r,K_0,\kappa,\lambda,\tau_c>0$, and the dependent variables $u=u(x,t)$, $c=c(x,t)$ are the cell density and chemoattractant concentration, respectively. $D_t^{\alpha,\lambda}$ is tempered fractional derivative notation.
\par 
Here $\mathcal{L}^{(\lambda)}_{\alpha}$ is the \emph{tempered fractional Laplacian}, e.g.
\begin{equation}\label{eq:laplacian_nonlocal}
\mathcal{L}^{(\lambda)}_{\alpha}u(x)
= C_{n,\alpha}\int_{\mathbb{R}^n}\frac{u(x)-u(y)}{|x-y|^{n+\alpha}}\,e^{-\lambda|x-y|}\,dy,
\end{equation}
where $C_{n,\alpha}$ is a constant, and this reduces to the standard fractional Laplacian when $\lambda=0$. Indeed, the TFKS system provides a flexible bridge between tempered and fractional models.

We now apply the techniques presented in Section \ref{sec:priliminaries} to the TFKS system \eqref{eq:TFKS}.
\par 
Let
\begin{equation}\label{eq:gauge_transform}
v(x,t)=e^{\lambda t}u(x,t), \qquad w(x,t) = e^{\lambda t}c(x,t).
\end{equation}
Applying the transformation (\ref{eq:gauge_transform}) to (\ref{eq:identity}), the system (\ref{eq:TFKS}) becomes
\begin{equation}\label{eq:TFKS_B}
\begin{cases} 
&{}^{C}D_t^{\alpha} v= \mathcal{L}^{(\lambda)}_{\alpha}\,v - \chi\,e^{-\lambda t}\,\left(v_{x}w_x + v w_{xx}\right)+ r\,v - \frac{r}{K_0}e^{-\lambda t}v^2, \\
& \tau_c\,w_t = (\tau_c\lambda - \kappa) w + D_c\,w_{xx} + v ,
\end{cases}
\end{equation}
where ${}^{C}D_t^{\alpha}$ denote a Caputo fractional derivative of order $0<\alpha < 1$, and $\lambda\geq 0$.
\par 
Throughout this work we adopt the short-jump diffusion approximation, which assumes that the characteristic jump length is sufficiently small compared with the spatial scale of variation of the solution. Under this assumption, applying Lemma \ref{lemma:1} in one dimension yields a second-order local approximation of the nonlocal operator.
\begin{equation}\label{eq:Laplacian}
D\,\mathcal{L}^\alpha_\lambda u \approx D_{\mathrm{eff}}\,u_{xx} = D_{\mathrm{eff}}\, e^{-\lambda t} v_{xx},\qquad \lambda > 0,\,\,\, 0<\alpha < 1,
\end{equation}
Consequently, in one spatial dimension, the system (\ref{eq:TFKS_B}) is equivalent to
\begin{equation}\label{eq:TFKS_C}
\begin{cases} 
&\mathrm{Eq 1}:\quad  {}^{C}D_t^{\alpha} v= D_{\mathrm{eff}}\,v_{xx} - \chi\,e^{-\lambda t}\,\left(v_{x}w_x + v w_{xx}\right)+ r\,v - \frac{r}{K_0}e^{-\lambda t}v^2 \\
&\mathrm{Eq 2}:\quad  \tau_c\,w_t = (\tau_c\lambda - \kappa) w + D_c\,w_{xx} + v 
\end{cases}
\end{equation}
Therefore, the system \eqref{eq:TFKS} reduced to the local system \eqref{eq:TFKS_C}, which serves as the basis for the Lie symmetry analysis presented in the following sections.
\subsection{Lie Point Symmetries of the Reduced TFKS}
\par 
Consider the vector field
\begin{equation}\label{eq:infB}
Y' = \tau(t,x,v,w)\,\partial_t + \xi(t,x,v,w)\,\partial_x + \phi_v(t,x,v,w)\,\partial_v + \psi_w(t,x,v,w)\,\partial_w.
\end{equation}
If the vector field (\ref{eq:infB}) can generate a symmetry of the system (\ref{eq:TFKS_C}), then the prolongation of $Y'$ vanish both Eq1 and Eq2 in (\ref{eq:TFKS_C}).
\par 
The invariance condition for Eq2 in (\ref{eq:TFKS_C}) takes the form:
\begin{equation*}\label{eq:inf3}
\mathrm{pr}\,Y' \left[ \tau_c\,w_t - (\tau_c\lambda - \kappa) w - D_c\,w_{xx} - v \right] \Bigg|_{\mathcal{E}'} = 0,
\end{equation*}
This expands to:
\begin{equation}\label{eq:infB2}
\tau_c\,\psi_w^t - (\tau_c\lambda - \kappa) \psi_w - D_c\,\psi_w^{xx} - \phi_v = 0
\end{equation}
 where $\psi_w^t$ and $\psi_w^{xx}$ are the prolongations of $\psi_w$, and
\begin{eqnarray*}
&\psi_w^t &= \mathrm{D}_t(\psi_w) - w_t \mathrm{D}_t(\tau) - w_x \mathrm{D}_t(\xi),\nonumber\\
&\psi_w^{xx} &= \mathrm{D}_x^2(\psi_w) - w_x \mathrm{D}_x^2(\xi) - 2 w_{xx} \mathrm{D}_x(\xi) - w_t \mathrm{D}_x^2(\tau) - 2 w_{tx} \mathrm{D}_x(\tau) - w_{tt} \dots \nonumber
\end{eqnarray*}
where the operators $\mathrm{D}_j$ is the total derivative with respect to $j$. For simplicity in derivation, we make the common assumption that $\tau = \tau(t)$ and $\xi = \xi(x,t)$.
\par

Substitute $w_t$ from Eq2:
\begin{equation*}
w_t = \frac{1}{\tau_c} \left[ (\tau_c\lambda - \kappa) w + D_c\,w_{xx} + v \right].
\end{equation*}
\par 
Equating coefficients of derivatives in (\ref{eq:infB2}) to zero yields
\begin{eqnarray}
\label{eq:coefB1}
&&\xi = c_1 x + c_2(t),\;\; \xi_{xx}=0,\;\;\psi_w = A(x,t)w + B(x,t),\\
&&\phi_v = C(x,t)v + D(x,t),\;\;\tau=\tau(t).\nonumber
\end{eqnarray}
\par
The invariance condition for Eq1 in (\ref{eq:TFKS_C}) takes the form:
\begin{equation}\label{eq:inf4}
\mathrm{pr}Y' \left[ {}^{C}D_t^{\alpha} v - \text{RHS} \right] \Bigg|_{\mathcal{E}'} = 0 \implies \phi_v^{\alpha,t} = \mathrm{pr}Y'(\text{RHS}),
\end{equation}
Using the Caputo prolongation formula presented in \cite{Thomas2024}, we obtain the prolongation of the Caputo derivative under point symmetries with $\tau = \tau(t)$. 
\par 
For $0<\alpha<1$ (Caputo derivative):
\begin{equation}\label{eq:inf5}
\mathrm{pr}\,Y'\!\left({}^{C}D_t^{\alpha}v\right)
= {}^{C}D_t^{\alpha}(Q_v)+\alpha\,\tau_t\,{}^{C}D_t^{\alpha}v,\
\end{equation}
where
\begin{equation*}
Q_v=\phi-\tau\,v_t-\xi^i v_{x_i},\qquad
Q_w=\psi-\tau\,w_t-\xi^i w_{x_i},
\end{equation*}
are the characteristics.
%
\par 
Putting (\ref{eq:coefB1}) and (\ref{eq:inf5}) together, the determining equations for (\ref{eq:TFKS_B}) are reduced to:
\begin{eqnarray}\label{eq:determining_B}
&&\tau_x = \tau_v = \tau_w = 0 \quad\Rightarrow\quad \tau = \tau(t),\\
&&\xi^i_v = \xi^i_w = 0 \quad\Rightarrow\quad \xi^i = \xi^i(x,t),\nonumber\\
&&\phi_{vv}=0, \; \phi_{vw}=0, \; \phi_{ww}=0, \quad
\psi_{vv}=0, \; \psi_{vw}=0, \; \psi_{ww}=0 \nonumber\\
&&\psi_w = A(x,t)w + B(x,t),\quad\phi_v = C(x,t)v + D(x,t).\nonumber
\end{eqnarray}

\begin{remark} 
The original TFKS system (\ref{eq:TFKS}), as well as the transformed system (\ref{eq:TFKS_B}) contain the nonlocal tempered fractional Laplacian and therefore do not satisfy the assumptions underlying the classical Lie symmetry framework employed here. In particular, for $\lambda\neq0$, the exponential tempering destroys time-translation invariance, and the nonlocal kernel restricts the admissible continuous symmetries to Euclidean isometries of the spatial variable. Consequently, systems (\ref{eq:TFKS}) and (\ref{eq:TFKS_B}) are invariant under spatial translations and rotations; this is the one continuous symmetry that can be justified directly at the level of the nonlocal system, and it survives unchanged in $n$ dimensions as $\partial_x$ (respectively $\partial_{x_j}$, $ x_j\partial_{x_k}-x_k\partial_{x_j}$). In one spatial dimension, this reduces to the translation generator $X=\partial_x$. 
\label{remark:3.1}
\end{remark}



\par 
Next, we obtain the admitted generators in the generic parameter case, where all model parameters are present and fixed at arbitrary nonzero values, and then show the conditional enlargements that need extra relations among parameters to restore symmetries which were broken in the generic case.
\subsection*{1) Generic case ($\lambda>0$ and $\chi,r,\kappa>0$; all constants fixed)}
This case is the most typical situation where all model parameters are present and fixed at arbitrary nonzero values.
\par 
The explicit factor $e^{-\lambda t}$ in (Eq 1) break time translations and most scalings when parameters are held fixed. The chemotaxis term and the logistic nonlinearity break vertical shifts and scalings of $v,w$.
\par 
Working through the determining system (\ref{eq:determining_B}) and the mixed constraints coming from the $e^{-\lambda t}$ term, all coefficients except a constant spatial translation are forced to zero:
\begin{eqnarray}
&& \tau=0,\,\xi_t=0,\,\xi_{i}^i+\xi_{i}^j=0\nonumber\\
&& \phi=\psi\equiv 0.
\end{eqnarray}
Therefore, the infinitesimal generator in one-dimension is
\begin{equation*}
\,X_1=\partial_x.\,
\end{equation*}
The infinitesimal generators in n-dimensions are
\begin{equation*}
\,X_1=\partial_x,\qquad X_2=x_j\partial_k - x_k\partial_j.
\end{equation*}
\par  
The transformation back to the original variables $(u, c)$ via $v=e^{\lambda t}u$ and $w=e^{\lambda t}c$ does not alter this fundamental result. The continuous symmetry remains exclusively that of spatial translation, $\partial_x$.
\subsection*{2) Conditional enlargements (only if extra parameter relations hold)}
The gauge transformation preserves the explicit time-dependence inherent in the tempered operators, which restricts the symmetries. Additional symmetries survive only under restrictive conditions. The common cases are outlined below.


\begin{itemize}
\item[\textbf{(a)}] \textbf{Time translation (with no parameter action)}
The gauge-transformed system (\ref{eq:TFKS_B}) admits a time-translation symmetry $\partial_t$ only when the tempering parameter vanishes $\lambda= 0$. However, under the short-jump diffusion approximation (Lemma \ref{lemma:1}), where the nonlocal tempered Laplacian is replaced by a local effective diffusion $D_{\mathrm{eff}}\,v_{xx}$, a modified scaling-translation symmetry may emerge when both chemotaxis and logistic terms are absent:
\begin{itemize}
\item[$\bullet$]
If $\chi=0$ and $r=0$ (purely linear diffusion with no chemotaxis, no logistic), based on the determining equations (\ref{eq:determining_B}) and invariance condition (\ref{eq:inf4}) under these conditions, the system admits
\begin{equation*}
X_{t}=\partial_t-\frac{\lambda}{2}\,x\,\partial_x.\,
\end{equation*}
This generator combines time translation with a compensating spatial scaling that exactly cancels the exponential tempering effects in the local approximation. This symmetry is valid only within the short-jump regime where the local effective diffusion replaces the nonlocal operator.
\item[$\bullet$] 
If either $\chi\neq 0$ or $r\neq 0$, the nonlinear coupling and logistic terms break even this modified symmetry, and no time-translation-like generator exists--neither in the local approximation nor in the full nonlocal system.
\end{itemize}
\item[\textbf{(b)}] \textbf{Pure Time Translation} From Remark (\ref{remark:3.1}), pure time translation $\partial_t$ is admitted if and only if the tempering parameter vanishes ($\lambda=0$). This condition is both necessary and sufficient, regardless of the values of other parameters $\chi, r$, or $\kappa$. In this case, from the determining equations (\ref{eq:determining_B}), the system admits the abelian Lie algebra $\mathfrak g=\langle X_2=\partial_t,\; X_1=\partial_x\rangle$ in one-dimension. In $n$-dimension, the algebra extends to include rotations: 
\begin{equation*}
\mathfrak g=\langle X_2=\partial_t,\; X_1=\partial_x,\; X_3=x_j\partial_{x_k} - x_k\partial_{x_j}\rangle,
\end{equation*}
where $i = 1,\ldots ,n$ and $j,k$ range over spatial indices with $j < k$.
\item[\textbf{(c)}] \textbf{Diffusive scaling}
The system's structure, which combines a fractional-in-time derivative for one equation, a first-order-in-time derivative for the other, and a non-local tempered-fractional Laplacian operator $\mathcal{L}_\alpha^\lambda$, a joint $(x,t)$ scaling generally fails to keep both equations invariant, as scaling is generally broken. Therefore, no diffusive scaling is admitted for fixed parameters (both $\lambda>0$ and $\lambda=0$).
\item[\textbf{(d)}] \textbf{Vertical scalings/shifts of $v$ or $w$}
\begin{itemize}
\item[$\bullet$]
Logistic and coupling terms ($rv$, $v^2$, $-\kappa w$, $+v$ in Eq 2) forbid additive shifts $v\mapsto v+\text{const}$, $w\mapsto w+\text{const}$, and forbid multiplicative scalings unless parameters co-transform. Hence, none are admitted in the generic constant-parameter setting.
\end{itemize}
\end{itemize}
\section{Optimal system of one-dimensional subalgebras}\label{sec:OptimalSys}
\par 
In this section, we classify all the one-dimensional subalgebras into subsets of conjugate
subalgebras using the method presented in \cite{olver2000applications}. This method takes a general
element of the Lie algebra and simplify it by choosing proper adjoint transformation.
\par 
The finite adjoint action by the one-parameter subgroup generated by $X_i$ is
\begin{equation*}\label{eq:adj}
\mathrm{Ad}(\exp(\varepsilon X_i))=\exp(\varepsilon\,\operatorname{ad}_{X_i})=\sum_{m=0}^\infty \frac{\varepsilon^m}{m!} (\operatorname{ad}_{X_i})^m.
\end{equation*}
\par 
For small $\varepsilon$ we can use \cite{olver2000applications}:
\begin{equation}
\operatorname{Ad}(\exp(\varepsilon X_i))X_j = X_j -\varepsilon [X_i,X_j] + \tfrac{\varepsilon^2}{2}[X_i,[X_i,X_j]]- \cdots.
\end{equation}

\par 
\subsection*{I) Generic case ($\lambda>0$ and $\chi,r,\kappa>0$; all constants fixed)}
\par 
The basis for the Lie algebra is obtained in section \ref{sec:lie_symmetries}, which is: $X_1=\partial_x$. Lie bracket is zero, as this is a one-dimensional algebra.
\par 
The adjoint action of
 $X_1$
is trivial, as:
$\operatorname{ad}_{X_1}=0$. 
This means that every finite adjoint transformation is the identity, $\operatorname{Ad}(\exp(\varepsilon X_1)) = \mathrm{Id}$. Consequently, the optimal system of one-dimensional subalgebras has only a single conjugacy class, which is represented by the subalgebra 
\begin{equation*}
\{\langle \partial_x\rangle\}.
\end{equation*}
\subsection*{II)  $\lambda=0$ (after gauge) — $\mathfrak g=\operatorname{span}\{\partial_t,\partial_x\}$ (abelian 2D)}
The basis for the Lie algebra consists of two generators $X_1=\partial_t$, and $X_2=\partial_x$. The Commutators are $[X_i,X_j]=0$ for all $i,j$, therefore
\begin{equation*}
\operatorname{Ad}(\exp(\varepsilon X_i))=\mathrm{Id}\qquad(i=1,2).
\end{equation*}
Since the adjoint action is trivial, conjugacy is equivalent to linear equivalence up to a nonzero scalar. This implies that any one-dimensional subalgebra spanned by $a\partial_t+b\partial_x$ is only equivalent to itself, up to a nonzero scaling factor. A standard and convenient optimal list of one-dimensional subalgebras is:
\begin{equation*}
\{\ \langle\partial_x\rangle,\ \langle\partial_t\rangle,\ \langle\partial_t + a\partial_x\rangle\ (a\ge 0)\ \}.
\end{equation*}
This list is \textit{optimal} because the subalgebra $\langle\partial_t + a\partial_x\rangle$ with $a\ge0$ effectively parameterizes all other one-dimensional directions, considering overall sign and scaling.
\par 
\subsection*{III) $\chi=0,\ r=0$ (no chemotaxis, no logistic)}
In this case, Lie algebra generated by
\begin{equation*}
X_1=\partial_x,\qquad X_2=\partial_t - \frac{\lambda}{2}\,x\,\partial_x.
\end{equation*}
(So $\mathfrak g=\operatorname{span}\{X_1,X_2\}$.)
\par 
The vector field Lie bracket is
\begin{equation*}
[X_2,X_1] = \frac{\lambda}{2}\,\partial_x = \frac{\lambda}{2}\,X_1.
\end{equation*}
Equivalently $[X_2,X_1]=(\lambda/2)X_1$ (so $[X_1,X_2]=-(\lambda/2)X_1$). This is a two-dimensional solvable (non-abelian) algebra where $X_2$ acts diagonally on $X_1$.
\par
\subsubsection*{Adjoint (choose ordered basis $(X_1,X_2)$)}
 Matrix in basis $(X_1,X_2)$:
\begin{equation*}
  A^{(1)}=\operatorname{ad}_{X_1}=
  \begin{pmatrix}
  0 & -\tfrac{\lambda}{2}\\[4pt]
  0 & 0
  \end{pmatrix},\qquad   A^{(2)}=\operatorname{ad}_{X_2}=
  \begin{pmatrix}
  \tfrac{\lambda}{2} & 0\\[4pt]
  0 & 0
  \end{pmatrix}.
\end{equation*}
Finite adjoint matrices (matrix exponentials):
\begin{itemize}
\item $\operatorname{Ad}(\exp(\varepsilon X_2))=\exp(\varepsilon A^{(2)})=
  \begin{pmatrix}
  e^{\tfrac{\lambda}{2}\varepsilon} & 0\\[4pt]
  0 & 1
  \end{pmatrix}.$\\
    Action on a general element $v=a X_1+b X_2$ is
  $(a,b)^\top \mapsto (e^{\tfrac{\lambda}{2}\varepsilon}a,\ b)^\top$ — i.e. $X_1$ is scaled.
\item 
$\operatorname{Ad}(\exp(\delta X_1))=\exp(\delta A^{(1)})$. Since $A^{(1)}$ is nilpotent,
\begin{equation*}
 \exp(\delta A^{(1)})=
  \begin{pmatrix}
  1 & -\tfrac{\lambda}{2}\delta\\[4pt]
  0 & 1
  \end{pmatrix}.
\end{equation*}
\end{itemize}
  Action on $v=(a,b)^\top$ is $(a - \tfrac{\lambda}{2}\delta\,b,\ b)^\top$ — so $X_1$-coefficient can be shifted by an amount proportional to the $X_2$-coefficient.

Next, we will obtain the optimal system.
\par 
Let $v=a X_1+b X_2$ (not both zero). Use adjoint actions to simplify:
\begin{itemize}
\item 
If $b\neq0$: apply $\operatorname{Ad}(\exp(\delta X_1))$ with $\delta=\tfrac{2a}{\lambda b}$ to kill the $X_1$-component: the transformed vector becomes proportional to $X_2$. Hence any 1-D subalgebra with $b\neq0$ is conjugate to $\langle X_2\rangle$.
\item  If $b=0$: $v=a X_1$ and this is conjugate to $\langle X_1\rangle$.
\end{itemize}
Therefore an optimal system of one-dimensional subalgebras is
\begin{equation*}
\{\ \langle X_1\rangle = \langle\partial_x\rangle,\qquad \langle X_2\rangle = \langle\partial_t - \tfrac{\lambda}{2} x\partial_x\rangle\ \}.
\end{equation*}
Note that these two are not conjugate to each other when $\lambda\neq0$.
\par 
The results presented in Table \ref{tab:summary_symmetries} provide a concise summary of the admitted Lie algebras and the corresponding optimal systems under the various parameter assumptions considered in this work.
\section{Similarity Reductions}\label{sec:SimilaritySolutions}
\par 
In this section, using the optimal systems we found in section \ref{sec:OptimalSys}, we obtain symilarity solutions for the reduced system (\ref{eq:TFKS_C}). We will convert fractional equations to ordinary equations by applying the Fractional Complex Transform (FCT) method, as detailed in \cite{Ibrahim2012,Li2010}.  
\begin{description}
\item[(I)] For generic case, the corresponding characteristic equation is:
\begin{equation}\label{eq:charctris1}
\dfrac{dt}{0}=\dfrac{dx}{1}=\dfrac{dv}{0}=\dfrac{dw}{0}.
\end{equation}
Integrating (\ref{eq:charctris1}) yield the following new variable and functions
\begin{equation}\label{eq:ex1}
v=v(t),\qquad w=w(t)
\end{equation}
By substituting (\ref{eq:ex1}) into equations (\ref{eq:TFKS_C}), we obtain reduced equation:
\begin{equation}\label{eq:reduce1}
\begin{cases}
{}^{C}D_t^{\alpha} v(t) &= r\,v(t) - \frac{r}{K_0}\,e^{-\lambda t}\,v(t)^2,\\
\tau_c\,\dfrac{dw}{dt}(t) &= (\tau_c\lambda-\kappa)\,w(t) + v(t).
\end{cases}
\end{equation}
\par 
\item[ (II)] For $\lambda=0$ (no tempering after gauge),
\par 
\begin{itemize}
\item[\textbf{II.A)}] $\langle X_1\rangle=\langle\partial_x\rangle$: 
In this case, the invariant solutions are independent of the spatial variable, i.e.,  $v=v(t)$, $w=w(t)$. System (\ref{eq:TFKS_C}) reduces to a purely temporal fractional ODE system:
\begin{equation}\label{eq:reduce2.2}
\begin{cases}
{}^{C}D_t^{\alpha} v = r v - \frac{r}{K_0} v^2,\\
\tau_c w_t=( -\kappa)w + v.
\end{cases}
\end{equation}
\item[\textbf{II.B)}] $\langle X_2\rangle=\langle\partial_t\rangle$ -- steady states:
\par  
In this case, all time derivatives vanish, yielding steady-state (time-independent) solutions. The system (\ref{eq:TFKS_C}) therefore reduces to a spatial fractional differential system:
\begin{equation}\label{eq:5.5}
\begin{cases}
0 &= D_{\mathrm{eff}}\, v_{xx} - \chi\big(v_x w_x + v w_{xx}\big) + r v - \frac{r}{K_0} v^2,\\[4pt]
0 &= D_c\,w_{xx} - \kappa w + v.
\end{cases}
\end{equation}
purely spatial second-order system; Caputo derivative vanishes on steady states.
\item[\textbf{II.C)}] For $\langle\partial_t+ a\,\partial_x\rangle$, with a time-Caputo order $0<\alpha < 1$, set the FCT clock
\begin{equation*}
\tau \;=\; \frac{t^{\alpha}}{\Gamma(1+\alpha)},
\end{equation*}
and use the fractional travelling coordinate
\begin{equation*}
\zeta \;=\; x - a\,\tau \quad\bigl(\text{so }\; \zeta = x - a\,t^{\alpha}/\Gamma(1+\alpha)\bigr).
\end{equation*}
We assume the invariant (travelling-wave--type) forms 
\begin{equation*}
v(x,t)=V(\zeta),\quad w(x,t)=W(\zeta),
\end{equation*}
with $V\in C^1([0,\zeta_T])$, and $v(0)=v_0,$
the standard FCT mapping implies
\begin{equation*}
{}^{\mathrm C}D_t^{\alpha}\,V(\zeta(t)) \;=\; \frac{d}{d\tau}V(\zeta(\tau))
\;=\; -\,a\,V'(\zeta).
\end{equation*}
The system (\ref{eq:TFKS_C}) reduces to:
\begin{equation}\label{eq:reduced_II.C}
\begin{cases}
-a\,V' &= D_{\mathrm{eff}}\,v'' - \chi\bigl(V'W' + V W''\bigr) + rV - \frac{r}{K_0}V^2,\\
-\tau_c a\frac{\alpha t^{\alpha-1}}{\Gamma (1+\alpha)} W' &= (\tau_c \lambda -\kappa) W + D_c\,W'' + V,
\end{cases}
\end{equation}
with $t=\left( \frac{\Gamma (1+\alpha)(x-\zeta)}{a}\right)^\frac{1}{\alpha}$.
\begin{remark}
For $0<\alpha<1$ and $\lambda>0$, this system retains explicit $t$-dependence. Consequently, the travelling-wave reduction is not a closed system of ordinary differential equations in $\zeta$ for the general fractional-order tempered case.
\end{remark}
\end{itemize}
\item[\textbf{(III)}] Degenerate linear case ($\chi=0,\ r=0$):
In the absence of chemotaxis and logistic growth, the system becomes linear, allowing exact analytical reduction. We consider two admissible symmetry generators corresponding to the optimal system identified in Section \ref{sec:OptimalSys}.
\begin{itemize}
\item[\textbf{III.A)}]  $\langle X_1\rangle=\langle\partial_x\rangle$, in this case, invariant variables are $v=v(t),\ w=w(t)$. Substituting these invariant variables into (\ref{eq:TFKS_C}), we obtain the reduced linear system
\begin{equation}\label{eq:5.9}
\begin{cases}
{}^{C}D_t^{\alpha} v = 0,\\
\tau_c w_t = (\tau_c\lambda-\kappa) w + v,
\end{cases}
\end{equation}
Since $0<\alpha <1$, $^{\text{C}}D_t^{\alpha}v=0$ implies $v(t)\equiv v_0$ (constant). Substituting into the second equation in (\ref{eq:5.9}), and solving with $w(0)=w_0$ yields
\begin{equation*}
w(t)=\Big(w_0+\frac{v_0}{\kappa-\tau_c\lambda}\Big)\,e^{\frac{\lambda-\kappa}{\tau_c}t}-\frac{v_0}{\kappa-\tau_c\lambda},
\quad\text{for } \kappa\neq\tau_c\lambda,
\end{equation*}
and in the resonant case $\kappa=\tau_c\lambda$,
\begin{equation*}
w(t)=w_0+\frac{v_0}{\tau_c}\,t.
\end{equation*}
Thus, under spatial uniformity and vanishing nonlinearities, the solutions remain exponential (or linear in the resonant case), controlled solely by decay and tempering parameters.
\item[\textbf{III.B)}]  $\langle X_t\rangle=\langle \partial_t - \tfrac{\lambda}{2}x\partial_x\rangle$ — similarity scaling
\par
Using its characteristics, we obtain invariant variable as follows:
\begin{equation*}
\frac{dt}{1}=\frac{dx}{-{\frac{\lambda}{2}} x} \quad\Rightarrow\quad
\ln x = -\tfrac{\lambda}{2} t + \mathrm{const} ,
\end{equation*}
which defines the similarity coordinate
\begin{equation*}
\zeta = x\,e^{\lambda t/2}.
\end{equation*}
Letting new variables be $v(x,t)=V(\zeta),\ w(x,t)=W(\zeta)$.
\par
Computing derivatives ($d/d\zeta$), we obtain:
\begin{equation}\label{eq:reduced_III_B}
\begin{cases}
v_x = e^{\lambda t/2} V'(\zeta),\quad v_{xx}=e^{\lambda t} V''(\zeta),\\
v_t = \tfrac{\lambda}{2}\zeta V'(\zeta)\, +\, {\text{(history terms)}}
\end{cases}
\end{equation}
Indeed, $v(x,t)=V(\zeta(x,t))$, so $v_t=V'(\zeta)\partial_t\zeta$ and the Caputo derivative acts non-locally.
\par 
Substituting these expressions into system (\ref{eq:TFKS_C}), with $\chi=r=0$ yields, for the Eq 1,
\begin{equation*}\label{eq:reduced_III_B2}
\,{}^{C}D_t^{\alpha}\big[V(\zeta(t))\big] \;=\; D_{\mathrm{eff}}\,V''(\zeta)\,.
\end{equation*}
The left-hand side (LHS) remains the Caputo derivative with respect to time of the composite function $V(\zeta(t))$. This term is non-local in time and does not automatically become an ordinary differential equation (ODE) in $\zeta$ unless further assumptions are made about history or functional form of $V$.
\par 
In Eq 2, the term $w_t$ becomes $\tfrac{\lambda}{2}\zeta W'(\zeta)$. The $w_{xx}$ term yields $D_c w_{xx}=D_c e^{\lambda t} W''(\zeta)$. That factor $e^{\lambda t}$ generally reintroduces explicit $t$-dependence, spoiling a closed $\zeta$-system unless one of the following holds:
\begin{itemize}
\item[$\bullet$]
 $D_c=0$ (no diffusion in $w$); 
\item[$\bullet$]
$W$ is chosen to carry a time prefactor (i.e. allow $\psi_w\neq0$ so $w$ rescales under $X_t$); 
\item[$\bullet$] 
some special balance between parameters occurs.
\end{itemize}
If $D_c=0$, Eq2 reduces to
\begin{equation*}
\tau_c \frac{\lambda}{2}\zeta W'(\zeta) = (\tau_c\lambda-\kappa)W(\zeta) + V(\zeta).
\end{equation*}
\par 
In summary, for the generator  $\langle X_t\rangle$, Eq1 reduces cleanly on its diffusion RHS to $D_{\mathrm{eff}}\, V''(\zeta)$,However, the Caputo derivative on the LHS retains its non-local, time-dependent nature. Eq2 only closes in $\zeta$ under extra parameter restrictions (e.g. $D_c=0$) or with a nontrivial scaling for $w$.
\end{itemize}
\end{description}
\section{Recovering Solutions of the TFKS System}\label{sec:ExactSolutions}
\par 
In this section, we find some exact solution for system (\ref{eq:TFKS_C}) based on the reduced equations. To relate these results to the original tempered-fractional Keller--Segel model, we reconstruct the original variables by applying the inverse gauge transformation. Consequently, the obtained expressions represent analytical solutions of the approximated TFKS model and provide approximate descriptions of the dynamics of the original nonlocal system (\ref{eq:TFKS}) within the validity of the short-jump diffusion approximation.
  
\begin{description}
\item[(I)] \textbf{For generic case}\\
The reduced system (\ref{eq:reduce1}) is equivalent to a fractional system. Using the standard FCT change of variable
\begin{equation*}
\zeta=\frac{t^\alpha}{\Gamma(1+\alpha)} \qquad(0<\alpha\le1).
\end{equation*}
For sufficiently smooth $v$ one has the well-known mapping for the Caputo derivative under FCT:
\begin{equation*}
{}^{C}D_t^\alpha v(t)=\frac{d v}{d\zeta}(\zeta).
\end{equation*}
Apply this transform to the first equation of (\ref{eq:reduce1}). Denote $t$ as a function of $\zeta$ by
\begin{equation*}
t(\zeta)=\big(\Gamma(1+\alpha)\,\zeta\big)^{1/\alpha}.
\end{equation*}

The first equation in (\ref{eq:reduce1}) becomes the ordinary (in $\zeta$) ODE
\begin{equation*}
\frac{dv}{d\zeta}=r\,v - \frac{r}{K_0}\,e^{-\lambda t(\zeta)}\,v^2.
\end{equation*}
This is a Bernoulli equation (power $2$). Introduce $y(\zeta)=1/v(\zeta)$. Then
\begin{equation*}
\frac{dy}{d\zeta}+ r\,y = \frac{r}{K_0}\,e^{-\lambda t(\zeta)},\qquad y(0)=\frac{1}{v_0}.
\end{equation*}
This is linear in $y$, so solve by integrating factor. 
\par 
Integrating factor: $\mathrm{e}^{r\zeta}$. Hence
\begin{equation*}
y(\zeta)=\mathrm{e}^{-r\zeta}\left(\frac{1}{v_0}+\frac{r}{K_0}\int_{0}^{\zeta}\mathrm{e}^{r s}\,e^{-\lambda t(s)}\,ds\right),
\end{equation*}
where $t(s)=(\Gamma(1+\alpha)\,s)^{1/\alpha}$. Returning to $v$ and to the original time $t$ (recall $\zeta=t^{\alpha}/\Gamma(1+\alpha)$) gives the explicit formula
\begin{equation*}
v(t)=\left\{\; \mathrm{e}^{-r\,\frac{t^\alpha}{\Gamma(1+\alpha)}}\!\left(\frac{1}{v_0}+\frac{r}{K_0}\int_{0}^{\zeta}\mathrm{e}^{r s}\,e^{-\lambda(\Gamma(1+\alpha)\,s)^{1/\alpha}}\,ds\right)\right\}^{-1}.
\end{equation*}
From the gauge transformation $u=e^{-\lambda t}v$, $u$ is
\begin{equation}\label{eq:6_AIU2}
u(t)=e^{-\lambda t}\left\{\; \mathrm{e}^{-r\,\frac{t^\alpha}{\Gamma(1+\alpha)}}\!\left(\frac{1}{v_0}+\frac{r}{K_0}\int_{0}^{\zeta}\mathrm{e}^{r s}\,e^{-\lambda(\Gamma(1+\alpha)\,s)^{1/\alpha}}\,ds\right)\right\}^{-1}
\end{equation}
This is an explicit quadrature solution -- the only non-elementary part is the integral
\begin{equation*}
\displaystyle \int_{0}^{\zeta}\mathrm{e}^{r s}e^{-\lambda(\Gamma(1+\alpha)s)^{1/\alpha}}ds,
\end{equation*}
which is a well-defined quadrature for fixed parameters.
\par 
Let $a=\frac{\tau_c\lambda-\kappa}{\tau_c}$. Using the integrating factor $e^{-a\,t}$ gives the exact solution for the second equation in (\ref{eq:reduce1}), we obtain
\begin{equation}\label{eq:CaseI_c1}
w(t)=e^{a\, t}\left[w_0+\frac{1}{\tau_c}\int_{0}^{t} e^{-a s}\,v(s),ds\right],
\end{equation}
where $v(t)$ is the exact solution obtained in (\ref{eq:6_AIU2}). with $\zeta(t)=t^\alpha/\Gamma(1+\alpha)$ and $t(\sigma)=(\Gamma(1+\alpha)\sigma)^{1/\alpha}$,
\begin{equation}\label{eq:caseI_c2}
v(t)=\frac{e^{r\zeta(t)}}{\displaystyle \frac{1}{v_0}+\frac{r}{K_0}\int_0^{\zeta(t)}e^{r\sigma}\,e^{-\lambda(\Gamma(1+\alpha)\sigma)^{1/\alpha}}\,d\sigma}.
\end{equation}
Substituting (\ref{eq:caseI_c2}) into (\ref{eq:CaseI_c1}) yields the fully explicit (quadrature) expression for $w(t)$:
\begin{equation}\label{eq:6_w1}
w(t)=e^{\frac{\tau_c\lambda-\kappa}{\tau_c} t}\!\left[
w_0+\frac{1}{\tau_c}\int_0^{t}
\frac{e^{-\frac{\tau_c\lambda-\kappa}{\tau_c} s},e^{,r,\zeta(s)}}
{\displaystyle \frac{1}{v_0}+\frac{r}{K_0}\int_0^{\zeta(s)} e^{r\sigma},e^{-\lambda(\Gamma(1+\alpha)\sigma)^{1/\alpha}},d\sigma},ds
\right],
\quad \zeta(s)=\frac{s^\alpha}{\Gamma(1+\alpha)},
\end{equation}
which is the exact $w(t)$ in the generic $(\lambda>0)$ case.
\par 
Transforming Eq. (\ref{eq:6_w1}) back, the original chemoattractant $c(t)$ obtain
\begin{equation}\label{eq:caseI_c}
c(t)=e^{-\kappa t/\tau_c}\left[c_0+\frac{1}{\tau_c}\int_0^t e^{\kappa s/\tau_c}\,u(s)\,ds\right],\quad u(s)=e^{-\lambda s}v(s),
\end{equation}
\item[(II)]  \textbf{Untempered case $\lambda=0$}
\par 
 Since the gauge was trivial at $\lambda =0$, the original variables coincide with the reduced equations.
 \begin{equation*}
u(x)=v(x),\qquad c(x)=w(x).
\end{equation*}
 \begin{itemize}
 \item[\textbf{II.A)}] Here $v=v(t)$, $w=w(t)$. The reduced initial value problem (IVP) is
 \begin{equation*}
{}^{C}D_t^\alpha v=r\,v-\frac{r}{K_0}v^2,\qquad
\tau_c\,w'=-\kappa w+v,\qquad v(0)=v_0,\ w(0)=w_0.
\end{equation*}

Use the Fractional Complex Transform (FCT)
\begin{equation}
\zeta=\dfrac{t^\alpha}{\Gamma(1+\alpha)},
\end{equation}
which maps the Caputo derivative (for sufficiently smooth $v$) to an ordinary derivative:
\begin{equation*}
{}^{C}D_t^\alpha v(t)=\dfrac{d v}{d\zeta}.
\end{equation*}
With ($\lambda=0$) the first equation becomes the ordinary logistic ODE in ($\zeta$):
\begin{equation*}
\frac{dv}{d\zeta}=r v-\frac{r}{K_0}v^2 = r v\Big(1-\frac{v}{K_0}\Big).
\end{equation*}
Let $A=\dfrac{K_0-v_0}{v_0}$. Integrating gives
\begin{equation*}
\ln\frac{v}{K_0-v}= r\zeta + C \quad\Longrightarrow\quad
\frac{v}{K_0-v}= \frac{v_0}{K_0-v_0}\,e^{r\zeta}.
\end{equation*}
Solving for $v$ and returning to $t$ (recall $\zeta=t^\alpha/\Gamma(1+\alpha)$) yields the closed-form
\begin{equation*}
v(t)=\dfrac{K_0}{1 + \displaystyle\frac{K_0-v_0}{v_0}\,\exp\!\Big(-\dfrac{r\,t^\alpha}{\Gamma(1+\alpha)}\Big)},
\end{equation*}
The second equation is linear first-order in real time, Using integrating factor ($e^{\kappa t/\tau_c}$). The general solution is
\begin{equation}\label{eq:6_w2}
w(t)=e^{-\kappa t/\tau_c}\Bigg(w_0+\frac{1}{\tau_c}\int_0^t e^{\kappa s/\tau_c},v(s),ds\Bigg),
\end{equation}
Substitute the explicit $v(s)$ into (\ref{eq:6_w2}), we obtain:
\begin{equation}\label{eq:6IIA1}
w(t)=e^{-\kappa t/\tau_c}\Bigg(w_0+\frac{K_0}{\tau_c}\int_0^t
\frac{e^{\kappa s/\tau_c}}{1 + A\,e^{-\frac{r s^\alpha}{\Gamma(1+\alpha)}}}\,ds\Bigg),
\qquad A=\frac{K_0-v_0}{v_0}.
\end{equation}
The integral in (\ref{eq:6IIA1}) generally has no elementary closed form for arbitrary $\alpha$, but it is a well-defined quadrature. 
transforming back, we obtain:
\begin{equation}\label{eq:6_AIIU}
u(t)=\frac{K_0}{1 + \dfrac{K_0-v_0}{v_0}\exp\big(-\tfrac{r\,t^\alpha}{\Gamma(1+\alpha)}\big)},\qquad
c(t)=e^{-\kappa t/\tau_c}\Bigg(c_0+\frac{1}{\tau_c}\int_0^t e^{\kappa s/\tau_c}u(s)\,ds\Bigg),
\end{equation}
where $c$ is equal to the tempered formula Eq. (\ref{eq:caseI_c}), when $\lambda=0$.
 \begin{figure}[H]
  \centering
  \includegraphics[scale=0.5]{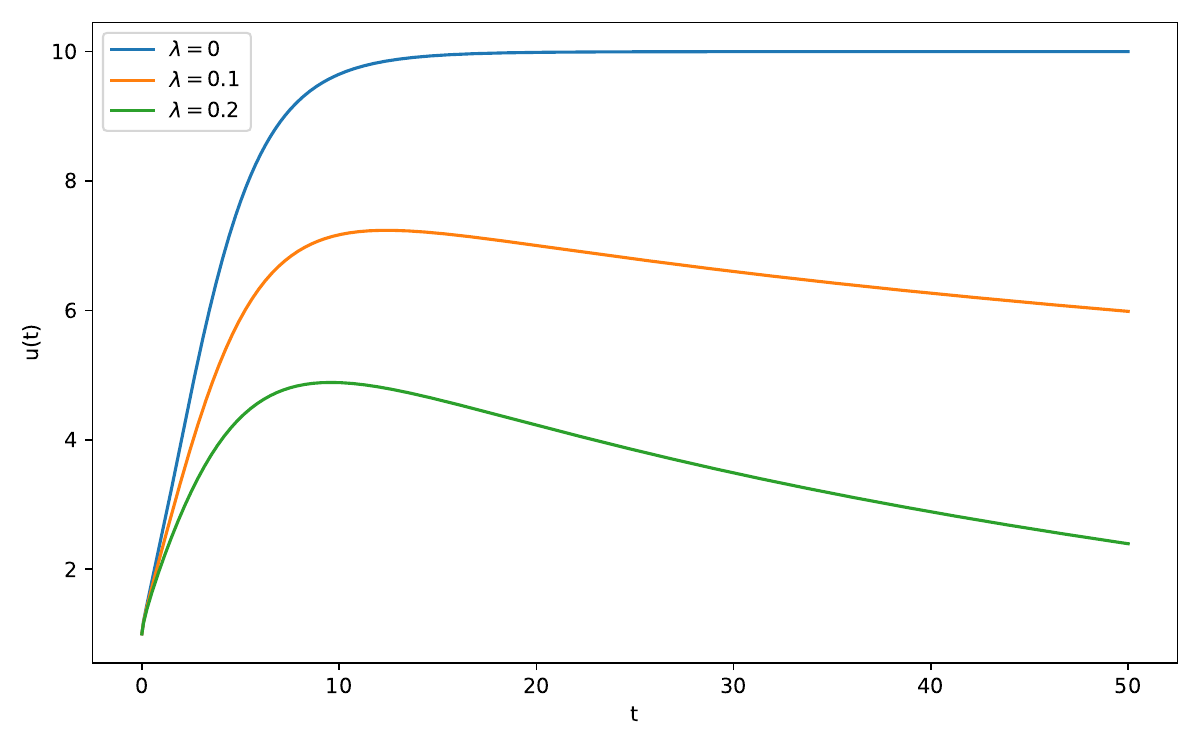}
  \caption{{\footnotesize Plot of $u(t)$ as given by Eq. (\ref{eq:6_AIU2}) for various values of the tempering parameter $\lambda$. Parameters $r = 1,\,K_0 = 10,\,v_0 = 1$.}}
\label{fig:AI2}
\end{figure}
Figure \ref{fig:AI2} highlights a clear contrast in the long-term behavior: while the untempered case ($\lambda=0$) exhibits sustained population survival, the introduction of tempering exponential exhibits a gradual monotonic decline toward extinction. Furthermore, the figure shows that increasing $\lambda$ both lowers the peak density attained during the transient phase and accelerates the subsequent decay.

\par 
The closed forms obtained in (\ref{eq:6_AIU2}) and (\ref{eq:6_AIIU}) allow us to establish the long-time behavior of the spatially homogeneous solutions analytically.

\begin{proposition}\label{proposition:2}
Let $0<\alpha\le 1$, $r,K_0>0$, and $0<v_0<K_0$. Then the solution (\ref{eq:6_AIIU}),
\begin{equation*}
u(t) = \frac{K_0}{1+A\,\exp\big(-\tfrac{r\,t^\alpha}{\Gamma(1+\alpha)}\big)}, \qquad A=\frac{K_0-v_0}{v_0}>0,
\end{equation*}
satisfies $v_0<u(t)<K_0$ for all $t>0$, is strictly increasing, and $u(t)\to K_0$ monotonically as $t\to\infty$.
\end{proposition}

\begin{proposition}\label{proposition:3}
Let $\lambda>0$, $0<\alpha<1$, and define
\begin{equation*}
J_\infty:=\int_0^\infty e^{rs}\,e^{-\lambda(\Gamma(1+\alpha)s)^{1/\alpha}}\,ds
\end{equation*}
Then:
\begin{itemize}
\item[(i)] $J_\infty<\infty$;

\item[(ii)] the solution (\ref{eq:6_AIU2})/(\ref{eq:caseI_c2}) satisfies, for all $t>0$,
$$\frac{e^{r\zeta(t)}}{M_\infty} \;\le\; v(t) \;\le\; v_0\,e^{r\zeta(t)}, \qquad \zeta(t)=\frac{t^\alpha}{\Gamma(1+\alpha)},\quad M_\infty:=\frac{1}{v_0}+\frac{r}{K_0}J_\infty,$$
so $v(t)\to\infty$ as $t\to\infty$;
\item[(iii)] the physical cell density $u(t)=e^{-\lambda t}v(t)$ satisfies
$$\frac{1}{M_\infty}\,\exp\!\Big(\frac{rt^\alpha}{\Gamma(1+\alpha)}-\lambda t\Big) \;\le\; u(t) \;\le\; v_0\,\exp\!\Big(\frac{rt^\alpha}{\Gamma(1+\alpha)}-\lambda t\Big),$$
and hence $u(t)\to 0$ as $t\to\infty$, for every $\lambda>0$.
\end{itemize}
\end{proposition}
\begin{proof}
\begin{itemize}
\item[] (i) Set $\beta=1/\alpha>1$ (since $0<\alpha<1$) and write the exponent of the integrand as $f(s)=rs-\lambda\Gamma(1+\alpha)^{\beta}s^{\beta}$. Since $\beta>1$, $s^{\beta-1}\to\infty$, so there exists $s_0$ such that $\lambda\beta\Gamma(1+\alpha)^{\beta}s^{\beta-1}\ge 2r$ for all $s\ge s_0$; on this range $f'(s)=r-\lambda\beta\Gamma(1+\alpha)^\beta s^{\beta-1}\le -r$, so $f(s)\le f(s_0)-r(s-s_0)$ for $s\ge s_0$. Thus $e^{f(s)}\le C e^{-rs}$ for $s\ge s_0$ and some constant $C$, which is integrable on $[s_0,\infty)$; combined with continuity of $e^{f(s)}$ on $[0,s_0]$, this gives $J_\infty<\infty$.

\item[] (ii) From (\ref{eq:6_AIU2}), $v(t)^{-1}=e^{-r\zeta}\big(\tfrac{1}{v_0}+\tfrac{r}{K_0}J(\zeta)\big)$, where $J(\zeta)=\int_0^\zeta e^{rs}e^{-\lambda t(s)}ds$ is nonnegative and nondecreasing in $\zeta$, with $J(\zeta)\le J_\infty$ for all $\zeta\ge0$ by (i). Hence the bracket lies in $[\tfrac1{v_0},M_\infty]$, and inverting gives the stated two-sided bound $e^{r\zeta}/M_\infty\le v(t)\le v_0 e^{r\zeta}$. Since $\zeta(t)\to\infty$ as $t\to\infty$, both bounds diverge, so $v(t)\to\infty$.

\item[] (iii) Multiply the bound in (ii) by $e^{-\lambda t}$ to obtain the stated bounds on $u(t)$. Because $0<\alpha<1$, $t^\alpha=o(t)$ as $t\to\infty$, so the exponent $rt^\alpha/\Gamma(1+\alpha)-\lambda t\to-\infty$ for every $\lambda>0$; both the upper and lower bound therefore tend to $0$, and the squeeze theorem gives $u(t)\to 0$. $\blacksquare$

\end{itemize}
\end{proof}

\begin{corollary}
Propositions \ref{proposition:2} and \ref{proposition:3} show that the distinction introduced by tempering is between \textit{survival at carrying capacity} ($\lambda=0$: $u(t)\to K_0$) and \textit{eventual extinction} ($\lambda>0$: $u(t)\to0$, after a finite transient rise).
\end{corollary}

\item[\textbf{II.B)}] 
Next, we obtain exact solutions in three possible cases.
\begin{itemize}
\item[1)] \textbf{Constant equilibria (global steady states on ($\mathbb{R}$) or bounded domains)}
\par 
 Set ($v\equiv V^*$) (constant). For constant $V^*$. Then the second ODE in (\ref{eq:5.5}) is linear, and its general solution will be
\begin{equation*}
w(x)=\frac{V^*}{\kappa} + A\,e^{\mu x} + B\,e^{-\mu x},\qquad \mu:=\sqrt{\kappa/D_c}.
\end{equation*}
Substituting $v\equiv V^*$ into the first ODE in (\ref{eq:5.5}),
\begin{equation*}
0 = rV^* - \frac{r}{K_0}(V^*)^2 \;\Longrightarrow\; V^*\in\{0,\,K_0\}.
\end{equation*}
Therefore the exact constant steady states (in the original variables $u,c$) are
\begin{equation}\label{eq:6II_B}
(u,c) \equiv (0,0), \qquad (u,c) \equiv \big(K_0,\; K_0/\kappa\big),
\end{equation}
and, on a finite interval, the paired family
\begin{equation*}
u(x)\equiv V^*,\qquad c(x)=\frac{V^*}{\kappa}+A\,e^{\mu x}+B\,e^{-\mu x},
\end{equation*}
with $V^*\in\{0,K_0\}$ and choosing $A,B$ to match boundary conditions (e.g., boundedness $\Rightarrow A=B=0$ on $\mathbb{R}$).
\item[2)] \textbf{Green's function (exact) representation for $c$ given any $u$}
\par 
For completeness (and for nonconstant $u$), the second ODE solves exactly as
\begin{equation}\label{eq:exact_IIB2}
c(x)=\big(G * u\big)(x)\;:=\;\int_{\mathbb{R}} G(x-\xi)\,u(\xi)\,d\xi,
\quad
G(z)=\frac{1}{2\sqrt{\kappa D_c}}\;e^{-\sqrt{\kappa/D_c}\,|z|}.
\end{equation}
Substitute this $c$ into the first equation of (\ref{eq:5.5}) to obtain an exact, closed integro-differential equation for $u$:
\begin{equation*}
0 = D_{\mathrm{eff}}u'' - \chi\Big(u'\,c' + u\,c''\Big) + r\,u - \frac{r}{K_0}u^2,
\quad c=G*u.
\end{equation*}
This characterizes all steady solutions (not only constants); solving it beyond the constant equilibria generally requires boundary conditions and, typically, analysis or numerics.

\item[3)] \textbf{Linear response around a constant state (still exact for the $c$--equation)}
\par 
If $u(x)\equiv V^*$ (with $V^*=0$ or $K_0$), then the $c$-equation reduces to the inhomogeneous Helmholtz ODE
\begin{equation*}
D_c c'' -\kappa\,c+u=0,
\end{equation*}
 with exact solution (\ref{eq:exact_IIB2}) on $\mathbb{R}$, boundedness forces $c\equiv V^*/\kappa$. 
\end{itemize}
\item[\textbf{II.C)}] Using the FCT travelling frame $\zeta = x - a\,t^{\alpha}/\Gamma(1+\alpha)$, with the ansatz $ v(x,t) = V(\zeta)$, we transform equation (\ref{eq:reduced_II.C}).
\par 
Solutions include:
\begin{itemize}
\item[\textbf{(i)}] \textbf{constant states $V^*\in\{0,K_0\}$ with explicit $W$}
Seek $V(\zeta)\equiv V^*$ (constant). Then $V'=V''=0$. The first ODE reduces to
\begin{equation*}
rV^* - \frac{r}{K_0}(V^*)^2 \;-\; \chi\,V^* W'' \;=\;0.
\end{equation*}
A sufficient way to satisfy this while keeping $V^*\neq0$ is to take $W''\equiv 0$ (i.e., $W$ affine). Then
\begin{equation*}
rV^* - \frac{r}{K_0}(V^*)^2 = 0 \quad\Rightarrow\quad V^*\in\{0,\;K_0\}.
\end{equation*}
The second ODE becomes a first-order linear ODE for $W$:
\begin{equation*}
-\,\tau_c a\, W' \;=\; -\kappa W + V^* \quad\Longrightarrow\quad
W(\zeta) \;=\; \frac{V^*}{\kappa} \;+\; C\,e^{(\kappa/(\tau_c a))\,\zeta}.
\end{equation*}
Thus the exact constant solutions (in the original variables $u,c$) are
\begin{equation*}
u(x,t)\equiv V^*\in\{0,K_0\},\qquad
c(x,t)= W(\zeta)\;=\;\frac{V^*}{\kappa} + C\,\exp\!\Bigl(\frac{\kappa}{\tau_c a}\,\bigl[x-a\,t^{\alpha}/\Gamma(1+\alpha)\bigr]\Bigr).
\end{equation*}
\item[\textbf{(ii)}] \textbf{$\chi=0$ Fisher--KPP--type profile }
With $\chi=0$, the first ODE is the nonlocal travelling-wave Fisher--KPP equation with speed $a$ in $\zeta$.
\begin{equation*}
-a\,V' \;=\; D_{\mathrm{eff}}V'' + rV - \frac{r}{K_0}V^2, 
\end{equation*}
with endstates $V(-\infty)=K_0$, $V(+\infty)=0$. We chose these boundary conditions because they describe a propagating front moving with speed $a > 0$. In the context of the TFKS model, they describe a self-similar propagating front connecting two stable states: (i) a populated state ($V = K_0$, saturated density) behind the front; (ii) an unpopulated state ($V = 0$, no cells) ahead of the front.
\par 
%
\par 
Once $V$ is found (analytically or numerically), the $W$-equation can be obtained from the linear ODE:
\begin{equation*}
D_c\,W''+\,\tau_c a\,W' -\,\kappa W +  V(\zeta)=0\quad \Longrightarrow\quad 
W(\zeta) = \bigl(G * V\bigr)(\zeta) + A e^{r_1\zeta} + B e^{r_2\zeta},
\end{equation*}
where
\begin{equation*}
r_{1,2}=\frac{-\tau_c a \pm \sqrt{\tau_c^2 a^2 + 4\kappa D_c}}{2D_c},\qquad
G(\zeta)=\frac{1}{\sqrt{\tau_c^2 a^2 + 4\kappa D_c}}\,
\begin{cases}
e^{r_1\zeta}, & \zeta>0,\\[2pt]
-\,e^{r_2\zeta}, & \zeta<0,
\end{cases}
\end{equation*}
and $A,B$ are fixed by boundary conditions. 
\end{itemize}

\end{itemize}
 \item[(III)] \textbf{Degenerate linear case ($\chi=0,\ r=0$)}
\begin{itemize}
\item[\textbf{III.A)}]
\par 
Now, undo the gauge (\ref{eq:gauge_transform}): $u=e^{-\lambda t}v;\, c=e^{-\lambda t}w$ in Equations (\ref{eq:5.9}). Using $v_0=v(0)=u(0)=u_0$ and $w_0=w(0)=c(0)=c_0$ (since the gauge is identity at $t=0$), we obtain an exact solution of the original TFKS system (1.1) in the spatially uniform, linear case $(\chi=0,;r=0)$:
\begin{itemize}
\item[$\bullet$] \textbf{Generic case $\kappa\neq \tau_c\lambda$}
\begin{equation}\label{eq:6_IIIA1}
\,u(x,t)=u_0\,e^{-\lambda t},\qquad c(x,t)=\Big(c_0+\frac{u_0}{\kappa-\tau_c\lambda}\Big)e^{-\kappa t/\tau_c}-\frac{u_0}{\kappa-\tau_c\lambda}\,e^{-\lambda t}
\end{equation}
\item[$\bullet$] \textbf{Resonant case $\kappa=\tau_c\lambda$}
\begin{equation}\label{eq:6_IIIA2}
u(x,t)=u_0\,e^{-\lambda t},\qquad c(x,t)=e^{-\lambda t}\Big(c_0+\frac{u_0}{\tau_c}\,t\Big)
\end{equation}
\end{itemize}
These formulas are the exact space-independent solutions to (\ref{eq:TFKS_C}) under the reduction leading to (\ref{eq:5.9}) (i.e., $\chi=0,; r=0$, and invariance under $\partial_x$), with the Caputo order $0<\alpha< 1$ as introduced in the gauge framework. 

\item[\textbf{III.B)}]
We proceed to find exact solution of the reduced equation (\ref{eq:reduced_III_B}), which arose from the scaling generator $\langle\partial_{t} - \frac{\lambda}{2}x\partial_{x}\rangle$.
\par 

Under the local diffusion approximation (Lemma \ref{lemma:1}), the TFKS system reduces to
\begin{equation*}
D^{\alpha,\lambda}_t u = D_{\mathrm{eff}}\,u_{xx},\qquad 
\tau_c\,c_t = D_c\,c_{xx}-\kappa\,c + u,
\quad 0<\alpha< 1,
\end{equation*}
with the tempered Caputo identity $D^{\alpha,\lambda}_t u = e^{-\lambda t} \,{}^C\!D_t^\alpha\!\big(e^{\lambda t}u\big)$.
\par 
Gauge away tempering via $v(x,t)=e^{\lambda t}u(x,t)$. Then
\begin{equation*}
{}^C\!D_t^\alpha v=D_{\mathrm{eff}}\,v_{xx}.
\end{equation*}
Defining the operational time transform
\begin{equation*}
s=\frac{t^\alpha}{\Gamma(1+\alpha)}.
\end{equation*}
Under this transformation, the Caputo derivative maps to a first-order derivative with respect to $s$
\begin{equation*}
{}^C\!D_t^\alpha v=\frac{\partial v}{\partial s},
\end{equation*}
This converts the fractional PDE into the classical integer-order heat equation in $(s,x)$:
\begin{equation*}
\frac{\partial v}{\partial s}=D_\mathrm{eff}\, v_{xx},\qquad v(0,x)=v_0(x)=u_0(x).
\end{equation*}
The solution $v(s,x)$ is found by convolving the initial condition $u_0(x)$ with the Gaussian heat kernel $G_{D_\mathrm{eff}}(s,x)$:
\begin{equation*}
v(s,x)=(G_{D_\mathrm{eff}}(s)\,*\, u_0)(x)=\int_\mathbb{R} G_{D_\mathrm{eff}} (s,x-\xi)u_0(\xi)d\xi,
\end{equation*}
where
\begin{equation*}
G_{D_{\mathrm{eff}}}(s,x)=\frac{1}{\sqrt{4\pi D_{\mathrm{eff}}\, s}}\exp\!\Big(-\frac{x^2}{4D_{\mathrm{eff}}\,s}\Big).
\end{equation*}
Transforming back using $u(x,t)=e^{-\lambda t}\,v(s(t),x)$ gives the final exact solution for the cell density
\begin{equation*}\label{eq:IIIB_u}
u(x,t)=e^{-\lambda t}\,\big(G_{D_{\mathrm{eff}}}(s(t))\!*u_0\big)(x),
\qquad s(t)=\frac{t^\alpha}{\Gamma(1+\alpha)}\; 
\end{equation*}
For a point mass initial condition $u_0=M\delta$ yields
\begin{equation}\label{eq:IIIB_u_example}
u(x,t)=e^{-\lambda t}\,\frac{M}{\sqrt{4\pi D_{\mathrm{eff}}\,s(t)}}\,
\exp\!\Big(-\frac{x^2}{4D_{\mathrm{eff}}\,s(t)}\Big),
\end{equation}
which is the fractional tempered heat kernel (Gaussian in $x$ with fractional operational time $s(t)$ and overall tempering factor $e^{-\lambda t}$). It serves as a fundamental solution (Green's function) for this class of non-local PDEs. Any other linear problem can be solved by convolving the initial and boundary conditions with this kernel.
\par 
Next, we obtain an integral representation for $c$. The $c$-equation is first-order in real time $t$. Using Duhamel's Principle with spatial heat kernel for $D_c$ and exponential temporal weight:
\begin{equation}\label{eq:IIIB_c}
c(x,t)=e^{-\frac{\kappa}{\tau_c}t}\,\big(G_{D_c/\tau_c}(t)\!*c_0\big)(x)
\;+\;\frac{1}{\tau_c}\int_0^t e^{-\frac{\kappa}{\tau_c}(t-s)}\,
\big(G_{D_c/\tau_c}(t-s)\!*u(s,\cdot)\big)(x)\,ds
\end{equation}
where $G_{D_c/\tau_c}(x,t)=\dfrac{1}{\sqrt{4\pi D_c t}}\exp\!\big(-x^2/(4D_c t)\big)$.
 The first term shows the decay and spread of the initial concentration $c_0(x)$ in the absence of cells, which governed by its own decay rate $\frac{\kappa}{tau_c}$
 and diffusion $D_c.$ The second (integral) term shows the concentration driven entirely by the cell population. 
 Since $u(\tau ,x)$ decays exponentially due to tempering, this integral term confirms that the chemoattractant concentration will also decay exponentially in the long term, preventing unbounded accumulation of the chemical.
\end{itemize}
\end{description}

 \begin{figure}[H]
  \centering
  \includegraphics[scale=0.45]{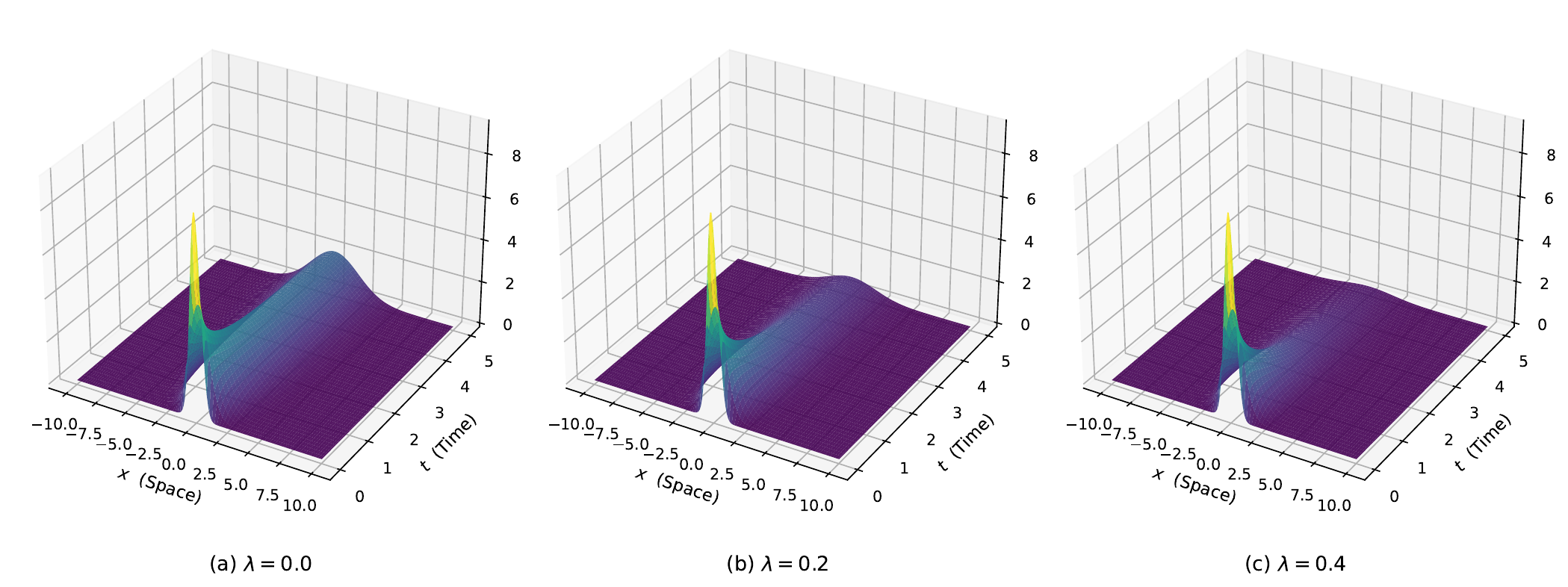}
 \caption{{\footnotesize Comparative three-dimensional visualization of the solution given by Eq. (\ref{eq:IIIB_u_example}) for three values of $\lambda$: (a) $\lambda=0$, (b) $\lambda = 0.2$, (c) $\lambda = 0.4$. Parameters: $\alpha=0.5,\, M=10,\,D_{\mathrm{eff}}=0.5$.}}
\label{fig:3dplot_lambda}
\end{figure}

\par 
Fig. \ref{fig:3dplot_lambda} illustrates how tempering prevents singularities, yielding solutions that remain bounded while preserving the fundamental subdiffusive character ($\alpha < 1$) of the spatial spreading. As $\lambda$ increases (Fig. \ref{fig:3dplot_lambda}, panels (b) and (c)), the exponential factor $e^{-\lambda t}$ introduces a biologically meaningful mortality or degradation mechanism acting uniformly on the cell population. This figure also shows that tempering progressively reduces population density without altering the fundamental spatial diffusion pattern. 


 \begin{figure}[H]
  \centering
  \includegraphics[scale=0.39]{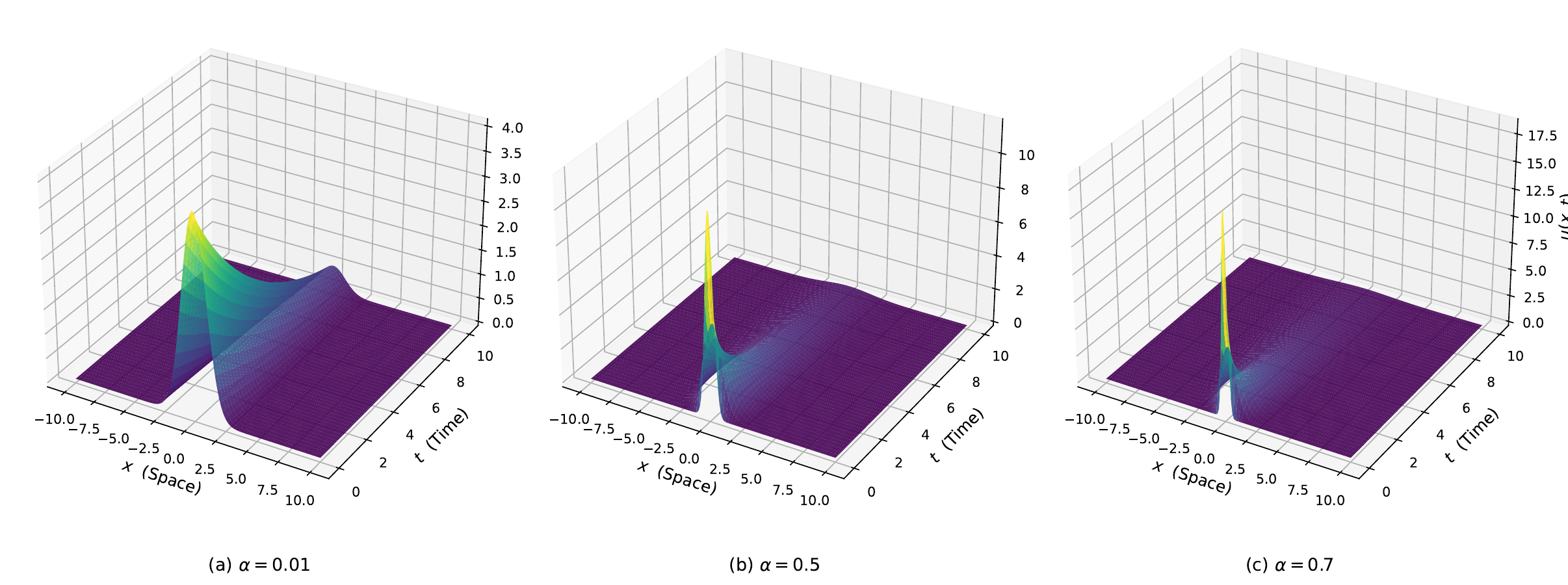}
 \caption{{\footnotesize Comparative three-dimensional visualization of the solution given by Eq. (\ref{eq:IIIB_u_example}) for three values of the fractional order: (a) $\alpha=0.01$, (b) $\alpha = 0.5$, (c) $\alpha = 0.7$. Parameters: $\lambda=0.2,\, M=10,\,D_{\mathrm{eff}}=0.5$.}}
\label{fig:3dplot_alpha}
\end{figure}

\par 
Fig. \ref{fig:3dplot_alpha} shows the influence of the fractional order $\alpha$ on the solution behavior. As $\alpha$ decreases, the operational time $s(t) = t^\alpha/\Gamma(1+\alpha)$ varies increasingly slowly with $t$, reflecting the subdiffusive nature of the process. Conversely, as $\alpha$ increases toward the classical diffusive limit, $s(t)$ grows more rapidly, yielding a sharp, high-amplitude peak that disperses quickly (panel (c)), with panel (b) marking an intermediate transition. Biologically, $\alpha$ reflects the degree of memory in individual cell trajectories: small $\alpha$ corresponds to strongly subdiffusive, trap-dominated motility -- consistent with cells constrained by adhesion or a dense extracellular matrix -- sustaining long-lived, spatially broad aggregates, whereas $\alpha$ closer to $1$ reflects faster, memoryless dispersal that produces sharp but transient concentration peaks.

\subsection{Biological Significance of Solutions}
\par 
The exact solutions provide key insights into the system's dynamics. Propositions \ref{proposition:2} and \ref{proposition:3} establish the precise long-time fate of the spatially homogeneous branch: Tempering shifts the outcome from \textit{survival at carrying capacity} ($\lambda=0$: $u(t)\to K_0$, Proposition \ref{proposition:2}) to \textit{eventual extinction} ($\lambda>0$: $u(t)\to0$ after a transient rise, Proposition \ref{proposition:3}). Biologically, this reveals the role of tempering: erodes the population's ability to sustain itself at a positive equilibrium, since the exponential tempering factor $e^{-\lambda t}$ in the governing operator acts as an effective time-dependent loss term that eventually overwhelms the logistic growth. This is consistent with the general biological expectation that populations are limited by resource exhaustion or mortality.
\par 
Eq. (\ref{eq:6II_B}) represents two possible long-term outcomes for the population in the spatially inhomogeneous steady-state analysis of Section \ref{sec:SimilaritySolutions}(II.B). The trivial state $(u,c)=(0,0)$ corresponds to extinction -- no cells, no chemoattractant concentration. The non-trivial state $(K_0, K_0/\kappa)$ represents a population at its environmental carrying capacity, where the chemoattractant level reflects a balance between production by cells and natural degradation. Proposition \ref{proposition:3} shows that, along the spatially homogeneous branch, the extinction state $(0,0)$ is in fact the generic long-time attractor whenever $\lambda>0$, while the carrying-capacity state is attained only in the untempered limit $\lambda=0$ (Proposition \ref{proposition:2}).
\par 
The temporal decay structure in Eq. (\ref{eq:6_IIIA1}) and Eq. (\ref{eq:6_IIIA2}) embodies the biological principle that unregulated exponential growth is incompatible with finite resources. Both cell density $u(x,t)$ and chemoattractant concentration $c(x,t)$ remain bounded and exhibit long-time decay in this degenerate linear regime ($\chi=0$, $r=0$), consistent with the extinction behavior established more generally in Proposition \ref{proposition:3}.
\par 
Eq. (\ref{eq:IIIB_u_example}) for $0<\alpha <1$ captures subdiffusion, where the mean squared displacement grows non-linearly with time $\langle x^2 \rangle \sim s(t) \sim t^\alpha $. As $t\rightarrow\infty$, the exponential decay $e^{-\lambda}$ dominates, causing the solution to decay exponentially to zero, which is consistent with Proposition \ref{proposition:3}.

\par 
The integral formulation in Eq. (\ref{eq:IIIB_c}) inherently demonstrates the non-local (history-dependent) nature of the concentration $c$. The value of $c$ at time $t$ depends on the entire history of cell density $u(\tau,x)$ from time $\tau=0$ to $t$. 
\section{Discussion and Future Research}
In this paper, we focus on the subdiffusion regimes with $0<\alpha< 1$ in time. A practical gauge framework to bring tempered Caputo derivatives to standard Caputo form, leading to a workable symmetry problem on a transformed system. The biological implications of our exact solutions suggest several directions for experimental validation. The tempering effect's role in shifting the asymptotic behavior from survival at the carrying capacity ($\lambda=0$) to eventual extinction ($\lambda>0$) warrants investigation through controlled experiments with bacterial chemotaxis systems.  However, several important theoretical and computational challenges remain for future investigation. For $1<\alpha<2$, the TFKS system exhibits fundamentally different behavior characterized by dynamic jumps with heavy-tail distributions. The mathematical structure in this regime presents additional complexities that our current gauge transformation approach cannot fully address. 
\par 

\section{Conclusion}
In this paper, we study the tempered-fractional Keller--Segel (TFKS) system, a model that accurately describes chemotaxis with anomalous, yet tempered, diffusion. In handling tempered kernels and Caputo derivatives, we built directly on the nonlocal Lie framework initiated by Gazizov and co-authors for fractional diffusion--adapting their symmetry perspective to operators whose kernels carry an exponential cutoff. Concretely, we used the tempered-Caputo identity and the gauge change $v=e^{\lambda t}u$, $w=e^{\lambda t}c$ to recast the TFKS system into one with standard Caputo time derivatives, making classical prolongation formulas applicable despite nonlocality. We obtained the optimal system of Lie algebras and used them to reduce the transformed TFKS system to a series of ordinary differential equations (ODEs) and found their corresponding exact solutions. Our results provide valuable insights into the long-term behavior and aggregation dynamics of the TFKS model.
\section{Statements and Declarations:}
This research received no external funding.
\section{Conflicts of Interest:}
The authors declare that they have no conflict of interest.

\clearpage
\begin{small}
\begin{table}[h!]
    \centering
    \begin{tabular}{|l|l|l|}
        \hline
        \textbf{Condition} & \textbf{Lie Algebra} &  \textbf{Optimal System} \\
        \hline
        Generic & $\mathrm{span}\{\partial_x\}$ & $\langle\partial_x\rangle$\\
        \hline
        $\lambda=0$ (untempered) & $\mathrm{span}\{\partial_t,\ \partial_x\}$ & $\{\langle\partial_x\rangle,\, \langle\partial_t\rangle,\,\langle\partial_t + a\partial_x\rangle\,(a\geq 0)\}$ \\
        \hline
        $\chi=0$, $r=0$ (no chemotaxis, no logistic) & $X_t=\partial_t-\frac{\lambda}{2}\,x\,\partial_x$ & $\{\langle\partial_x\rangle,\, \langle\partial_t- \frac{\lambda}{2}x\partial_x\rangle\}$ \\
        \hline
    \end{tabular}
    \caption{Lie algebra and optimal system under various parameter conditions.}
    \label{tab:summary_symmetries}
\end{table}
\end{small}

\end{document}